\documentclass[referee,a4paper,12pt,traditabstract]{swsc}

\usepackage{graphicx}
\usepackage{txfonts}
\usepackage{subfigure}

\usepackage{float}

\usepackage{epstopdf}
\usepackage[backref]{hyperref}
\usepackage{lineno}

\hypersetup{colorlinks=true,citecolor=cyan,urlcolor=cyan,linkcolor=blue}

\begin{document}

   \title{\Large{A Space weather information service} \\[-0.25cm]}

   \subtitle{\Large{based upon remote and in-situ measurements of coronal mass ejections heading for Earth}\\[1.2cm]
   \large{A concept mission consisting of six spacecraft in a heliocentric orbit at 0.72 AU}}

   \author{Birgit Ritter\inst{1}*
          \and
          Arjan J. H. Meskers\inst{2}
          \and
          Oscar Miles\inst{3}
          \and
          Michael Ru{\ss}wurm\inst{4}
          \and
          Stephen Scully\inst{5}
          \and
          Andr\'es Rold\'an\inst{6}
          \and
          Oliver Hartkorn\inst{7}
          \and
          Peter J\"ustel\inst{8}
          \and
          Victor R\'eville\inst{9}
          \and
          Sorina Lupu\inst{10}
          \and
          Alexis Ruffenach\inst{11}
          }

   \institute
            {German Aerospace Center, Institute of Aerospace Medicine, Radiation Biology Department, Linder H\"ohe, 51147 K\"oln, Germany,
            \email{\href{mailto:birgit.ritter@dlr.de}{birgit.ritter@dlr.de}}\\
         \and
            Delft University of Technology, Faculty of Mechanical Engineering, Mekelweg 2, 2628CD, Delft, The Netherlands\\
        \and
            University of Southampton, Faculty of Applied Sciences and Engineering, SO17 1BJ Southampton, United Kingdom\\
        \and
            Vienna University of Technology, Karlsplatz 13, Vienna, Austria\\
        \and
        	Department of Experimental Physics, National University of Ireland Maynooth, Co. Kildare, Ireland\\
        \and
            University of Granada, Faculty of Sciences, Avda. Severo Ochoa s/n, Granada, Spain\\
        \and
   			University of Cologne, Institute of Geophysics and Meteorology, Pohligstrasse 3, 50969 Cologne, Germany\\
        \and
            University of Stuttgart, Keplerstra{\ss}e 7, 70174 Stuttgart, Germany\\
        \and
            AIM Paris-Saclay Laboratory, CEA/Irfu Paris Diderot University CNRS/INSU, 91191 Gif-sur-Yvette, France\\
        \and
        	Politehnica University of Bucharest, Faculty of Electronics, Telecommunications and Information Technology, Splaiul Independentei 313, Romania \\
        \and
            Institut de Recherche en Astrophysique et Planetologie, Toulouse University (UPS), CNRS UMR 5277, France\\
             }

\abstract
{The Earth's magnetosphere is formed as a consequence of interaction between the planet's magnetic field and the solar wind, a continuous plasma stream from the Sun. A number of different solar wind phenomena have been studied over the past forty years with the intention of understanding and forecasting solar behavior. One of these phenomena in particular, Earth-bound interplanetary coronal mass ejections (CMEs), can significantly disturb the Earth's magnetosphere for a short time and cause geomagnetic storms. This publication presents a mission concept consisting of six spacecraft that are equally spaced in a heliocentric orbit at 0.72 AU. These spacecraft will monitor the plasma properties, the magnetic field's orientation and magnitude, and the 3D-propagation trajectory of CMEs heading for Earth. The primary objective of this mission is to increase space weather forecasting time by means of a near real-time information service, that is based upon in-situ and remote measurements of the aforementioned CME properties. The obtained data can additionally be used for updating scientific models. This update is the mission's secondary objective. In-situ measurements are performed using a Solar Wind Analyzer instrumentation package and flux gate magnetometers, while for remote measurements coronagraphs are employed. The proposed instruments originate from other space missions with the intention to reduce mission costs and to streamline the mission design process. Communication with the six identical spacecraft is realized via a deep space network consisting of six ground stations. They provide an information service that is in uninterrupted contact with the spacecraft, allowing for continuous space weather monitoring. A dedicated data processing center will handle all the data, and then forward the processed data to the SSA Space Weather Coordination Center which will, in turn, inform the general public through a space weather forecast. The data processing center will additionally archive the data for the scientific community. The proposed concept mission allows for major advances in space weather forecasting time and the scientific modelling of space weather.}


\titlerunning{A Space weather information service}

    \maketitle



\section{Introduction}
\label{Introduction}
The Earth and its near surroundings are affected by space weather, which is defined as "the physical and phenomenological state of natural space environments" (COST 724 final report \cite{COST}). The Sun is the main driver of space weather phenomena and coronal mass ejections (CMEs) are one of these phenomena. CMEs are created by plasma eruptions of solar material in a mass range between $10^{12}$ and $10^{13}$ kg (Meyer-Vernet \cite{meyer2007}, COST 724 final report \cite{COST}).
Magnetic reconnection, caused by the twisting and tangling of magnetic field lines, occurs in the solar corona, where a vast amount of energy is contained. This process results in an eruption of a magnetic field structure, potentially leading to a CME (Gla{\ss}meier and Scholer \cite{meier1991}; Gopalswamy \cite{gopalswamy2003}). However, at present, the trigger of a CME is not fully understood. The rate of occurrence of CMEs varies roughly according to the solar cycle, ranging from an average rate of 3 per day at solar maximum to an average rate of 1 per week at solar minimum (Meyer-Vernet \cite{meyer2007}). Earth-directed CMEs possibly impacting Earth's environment exhibit a minimal opening angle of 60 degrees and their travel velocities range from 50 to 3000 km/s (Howard et al. \cite{howard1985}; Gopalswamy et al. \cite{gopalswamy2004}; Gopalswamy \cite{gopalswamy2008}).
After its release, the CME plasma and magnetic field propagate through the interplanetary medium (IPM) and interact with the ambient solar wind. The manifestation of a CME in the IPM is sometimes called interplanetary coronal mass ejection (ICME), whereas the expression 'CME' is also accepted for both objects and is used throughout this article. The expansion of a CME is approximately self-similar within the IPM and its plasma density and magnetic field strength decrease faster than linearly during its propagation through the heliosphere (e.g. Gulisano et al. \cite{Gulisano2012}, and references therein). Moreover, comprehensive studies on CME propagation have been conducted by Manoharan (\cite{Manoharan2006}) and Manoharan (\cite{Manoharan2010}) using remote-sensing observations indicating that the magnetic energy associated with a CME is responsible for maintaining the CME structure during its propagation through the IPM.   
The shock front develops since the speed differential between the CME plasma and the usual solar wind plasma often exceeds the magnetosonic wave speed. It is followed by the shocked solar wind plasma, the sheath that is a turbulent region, and the driving ejecta of the CME.
\\The magnetic field and plasma properties of the driving ejecta are crucial for the process properties of the interaction between a CME and Earth's plasma environment.
As a matter of fact, CMEs can cause geomagnetic storms on Earth upon interaction with Earth's magnetosphere.

\noindent
Indeed, they are the primary cause for the most severe storms (Gosling et al. \cite{gosling1990}). Such storms are defined as intervals of time, in which the magnetospheric ring current is intensified as a result of increased energy and particle injections from solar wind plasma into Earth's magnetosphere-ionosphere system. The enhancement of the ring current generates perturbations of the geomagnetic field on the ground. Effects of geomagnetic storms on and near Earth include communication disruptions, current surges in power lines and radiation hazards to operating astronauts and spacecraft (e.g. COST 724 final report \cite{COST}, Reitz \cite{reitz2008}).
\\Coronagraphs, such as the Large Angle and Spectrometric COronagraph (LASCO, Brueckner et al. \cite{brueckner1995}) instrument onboard the Solar and Heliospheric Observatory (SOHO, Domingo et al. \cite{domingo1995}), are used to monitor the solar corona for disturbances such as CMEs (Hundhausen et al. \cite{hundhausen1984}). CMEs containing a dominant southward magnetic component are most geoeffective (Gonzales et al. \cite{gonzalez1994}), meaning they have a high ability to cause geomagnetic storms (Gopalswamy \cite{gopalswamy2008}). In this case, magnetic reconnection at the dayside of the magnetopause (Dungey \cite{dungey1961}) allows for energy of the ejecta to be transferred into the inner regions of the Earth's magnetosphere (Akasofu \cite{akasofu1981}), which enhances the ring current intensity. Besides having a specific magnetic field orientation, a CME's dimension and velocity are additional important factors that determine the effectiveness of a magnetic storm (Gosling et al. \cite{gosling1990}).
\\CME€™s can currently be detected in-situ by spacecraft at L1 (Lagrange-1 point), such as the Advanced Composition Explorer (ACE, Smith et al. \cite{Smith1998}), the Comprehensive Solar Wind Laboratory for Long-Term Solar Wind Measurements (WIND, Ogilvie et al. \cite{Ogilvie1995}), and SOHO. They allow for an effective warning time of about one hour before effects of a geomagnetic storm are detectable on Earth.
\\Previous mission proposals have considered how to increase this warning time. Examples include 'Geostorm' (West \cite{west1996}, \cite{west2004}), which is a solar-sail mission at 0.98 AU and 'Space Weather Diamond' (St.~Cyr et al. \cite{cyr2000}), which is a multi-spacraft mission at 0.9 AU.
'Geostorm' is supposed to be placed in a rearranged orbit at L1, which is shifted closer to the Sun due to the effect of a solar sail.
'Space Weather Diamond' consists of four spacecraft on eccentric heliocentric orbit seeming to circuit around the Earth 0.1 AU apart from it. All spacecraft are equally equipped with in-situ measurement instruments.

\noindent
This paper presents a new mission proposal to monitor CMEs. The mission constellation is comprised of six spacecraft placed in heliocentric circular orbits at 0.72 AU at a separation angle of 60 degrees from each other. All spacecraft are able to perform remote and in-situ measurements. The mission serves as a continuous information-service system, providing all necessary information of the heliosphere that allow for an increased space weather forecasting time and enhancement of scientific models regarding space weather. The data will help to forecast if there will be a magnetic storm and determine its severity, and will also provide information on radiation levels for astronauts and space missions further away from the Sun. This mission poses a large advancement in protecting human's health and technology, and in preventing technological crashes and negative results for Earth's eco-system.

\noindent
In contrast to Geostorm, the proposed mission is based on well-known and tested technical equipment, ready for the usage as a steadily operating system within the next decades.
'Space Weather Diamond' is called a monitoring system, as well as the here presented mission proposal. However, 'Space Weather Diamond' focuses on the reactor of the system, the Earth, whereas the here presented mission monitors the actual actor, the Sun, including remote measurements systems allowing a more comprehensive view on the state of space weather than single point measurements. This mission proposal sets the spacecraft significantly closer to the Sun, at 0.72 AU and thereby yields a much longer forecasting time, which is the most important advantage of this mission. 

\noindent
However, with an increase of forecasting time comes naturally a decrease of accuracy. Lindsay et al. (\cite{lindsay1999}) concluded that in-situ measurements between 1.0 AU and 0.7 AU in the equatorial plane within 10 degrees east to 5 degrees west of the Earth-Sun line allow substantial space weather forecasts. While the underlying model of this conclusion is a simple linear model by Burton et al. (\cite{burton1975}), the authors of the paper at hand believe that advanced data-driven CME models (e.g. Luhmann et al. \cite{luhmann2004}, T\'{o}th et al. \cite{toth2005}), calibrated by data of the early mission-phase, will allow for a reliable warning system, if at least one spacecraft is located within a range $\pm$30 degrees apart from the Earth-Sun line.
\\The proposal was originally entitled as the CARETAKER mission, and has been designed by a group of Bsc. and Msc. students, PhD-candidates and Postdocs from the ESA member states, as part of the Alpbach Summer School\footnote{http://www.summerschoolalpbach.at/} 2013. This summer school was organized and funded by an international cooperation between the European Space Agency (ESA), the Aeronautics and Space Agency of Austria (part of the Austrian Research Promotion Agency FFG), the International Space Science Institute (ISSI), and Austrospace, the association of Austrian space industries and research institutions.
\\In this paper the CARETAKER mission proposal is presented. The paper is structured as follows: in Section 2 the mission overview is given including the mission statement and requirements, and the operational concept. The flight segment is described in Section 3, followed by the communication segment in Section 4 and the operation and ground segment in Section 5. In Section 6 the data processing is discussed, and Section 7 covers the budgets of the mission followed by a risk analysis in Section 8, and a cost analysis in Section 9. The paper ends with a conclusion in Section 10.

\section{Mission Overview}
\label{Mission Overview}
The mission consists of six identical spacecraft in a heliocentric orbit at 0.72 AU, performing in-situ and remote measurements. The obtained data will be communicated periodically via a direct link between spacecraft and Earth through a dedicated deep-space network, while communication with the spacecraft located behind the Sun is naturally excluded. The deep-space network consists of six ground stations that are located in a way to ensure continuous contact with the spacecraft. Two ground stations will operate simultaneously, each alternating communication between three spacecraft. All obtained information will be collected by a single mission operations center that passes it along to a data processing center where the data will be processed and submitted to the SSA Space Weather Coordination Center\footnote{http://swe.ssa.esa.int/web/guest/ssa-space-weather-activities} (SSCC), and the information will additionally be archived for the scientific community. It will be the SSCC's responsibility to send out a warning to respective users requiring such an alert system in order to trigger their precautions that need to be undertaken.

\subsection{Mission Statement and Requirements}
\label{Mission Statement and Requirements}
The mission states that: \emph{the system consists of a near real-time information-service, based on the physical properties of solar Coronal Mass Ejections.}\\
The primary objective of the mission is to \emph{obtain information about CMEs as input for a CME warning system}, which covers information about:
\begin{enumerate}
\item the propagation trajectory of CMEs heading for Earth, and
\item the physical properties of CMEs.
\end{enumerate}
The secondary mission objective is to \emph{improve CME models by remote and in-situ multipoint measurements.}

\noindent
Achieving the above stated mission objectives requires that the mission shall:
\begin{enumerate}
  \item monitor the three dimensional trajectory of CMEs that head for Earth;
  \item measure the CME's magnetic field orientation and magnitude (related to geoeffectiveness);
  \item determine the CME's propagation envelope within an accuracy of less than 4.3 arcmin (i.e. the angular diameter of the Earth'€™s magnetosphere as viewed from the Sun);
  \item remotely monitor CMEs between 2 to 15 solar radii (for enhancement of scientific models, Thernisien et al. \cite{thernisien2009});
  \item provide a minimum forecasting time of 12 h, measured upon reception of in-situ measurement values;
  \item measure the following plasma properties: the 3D velocity distribution of protons and electrons, and the composition of heavy ions up to 56 amu/q, all measured with a time resolution of 60 s for the detection of the CME shock front (Richardson and Cane \cite{richardson2004});
  \item measure the low-energy ion particle flux in the range of 0.26 keV/q to 20 keV/q as well as the low-energy electron flux in the range of 1 eV to 5 keV;
  \item measure the magnetic field with a resolution of 0.1 nT in the range between -200~nT and 200~nT (Burlaga \cite{burlaga2001});
  \item provide the SSCC with data, processed according to their standards (allowing them to construct a space weather forecast for the general public);
  \item have an operational life time of 5 years (with a possible extension of 5 years);
  \end{enumerate}

\subsection{Operational Concept}
\label{Operational Concept}
The mission combines results from remote stereo images and multiple in-situ measurements in order to determine the trajectory and physical properties of CMEs. This is achieved by having six identical spacecraft in a heliocentric orbit at 0.72 AU, at a separation angle of 60 degrees apart. This separation angle is driven by the argument that CMEs with an angular extent larger than 60 degrees have a notable impact on Earth (Gopalswamy \cite{gopalswamy2008}). All spacecraft are equipped with a coronagraph (2-15 solar radii field of view, 5 min cadence), a Solar Wind Analyzer instrumentation package, and flux gate magnetometers, both sampling at 1 Hz. The in-situ measurements are performed continuously by all spacecraft, whereas only two spacecraft will be performing remote measurements simultaneously, limiting the amount of data  communicated back to Earth. The two spacecraft performing remote measurements will be the ones having the best point of view for stereoscopic imaging with respect to Earth.
The three spacecraft closest to the Earth (two of which perform remote sensing as well as in-situ) will downlink their data every 15 minutes whereas the three distant spacecraft (only making in-situ measurements) will downlink every 45 minutes. The processing of stereoscopic remote sensing of a CME will result in information regarding the CME's trajectory, velocity and size. These results are processed in a standardized format (in accordance with the SSCC) and will be passed on to the SSCC within approximately 45 minutes after the CME's occurrence at the Sun. The CMEs travelling towards Earth will be continuously monitored up to 15 solar radii. In-situ measurements of the solar wind and passing CMEs will be made at 0.72~AU. This information will also be processed according to standards and made available to the SSCC within 45 minutes after detection. From the moment on in which the in-situ measurement data has been processed it is possible to determine the geoeffectiveness of the approaching CME, resulting in a precaution time of at least 12 hours before the CME reaches Earth's magnetosphere (estimated for fast CMEs with a travelling speed of 1000 km/s when arriving at the Earth).

\section{Flight Segment}
\label{Flight Segment}

\subsection{Orbit}
\label{Orbit and Launch}
From the mission requirements it can be concluded that the CARETAKER concept must consist of six spacecraft at a heliocentric orbit of at most 0.8 AU, each with a 60 degrees separation. Analysis of the options for realizing this concept led to the optimal solution of inserting the spacecraft into a Venus orbit at 0.72 AU using gravity assist maneuvers (GAMs) at Venus to reduce the required propellant mass. The dependance on the GAM limits the launch window to a maximum of 3 weeks every 19 months.\\
Solar cycles 25 and 26 (counting began in 1761, on cycle after the end of the Maunder minimum) will likely occur between 2020-2031 and 2031-2042, respectively (DeRosa et al. \cite{derosa2012}). The maximums of those cycles will thus be around 2026 and 2037, with an enhanced number of solar events expected to occur. The first spacecraft will be in final orbit in 2027, the last in 2030. Thus during calibration of the payload, some spacecraft are likely to be exposed to extreme events. The nominal mission will cover an increasing solar activity during the 5 years of operations, even if the launch is delayed. With the extended mission lifetime, an entire solar cycle can be monitored.

\subsection{Launch}
\label{Launch}
A required characteristic energy of 5.71 km$^2$/s$^2$ was calculated to insert the spacecraft into the transfer orbit. The calculation is based on the mass of the spacecraft and its overall trajectory. Table~\ref{tab:launchertradeoff} displays the resulting launcher options investigated for this mission.

\begin{table}[h!]
\centering
\caption{Launcher tradeoff between Soyuz and Ariane 5. Despite the higher costs, Ariane 5 launchers are considered the better option with the Soyuz being a viable descoping option.}
\begin{tabular}{lccc}
Parameter \verb+\+ Launcher && Soyuz & Ariane 5 \\
\hline \hline
Performance at Required Escape Speed && 1850 kg & 5255 kg\\
Cost per Launch && 75 M EUR  & 160 M EUR \\ 
Cylindrical Fairing Dimensions (h x $\varnothing$) && 5060 x 3860 mm\textsuperscript{2} & 10039 x 4570 mm\textsuperscript{2}\\
Number of Launchers Required && 3 & 2 \\
Total Mass Delivered to Transfer Orbit && 5550 kg & 5255 kg\\
\hline
Total Cost && 225 M EUR & 320 M EUR \\ 
\hline
\end{tabular}
\label{tab:launchertradeoff}
\end{table}

\noindent
As the launch window constraint has more significant implications for the Soyuz launch, it was concluded that the delay on mission operations was severe enough to justify using two Ariane~5 launchers. The Soyuz launcher is however a viable descoping option.\\
Kourou in French Guiana is chosen as the designated spaceport because of its proximity to the equator.

\subsection{Orbital Insertion Procedure}
\label{Orbital Insertion Procedure}
\begin{figure}[hbtp]
   \centering
\includegraphics[width=0.8\textwidth]{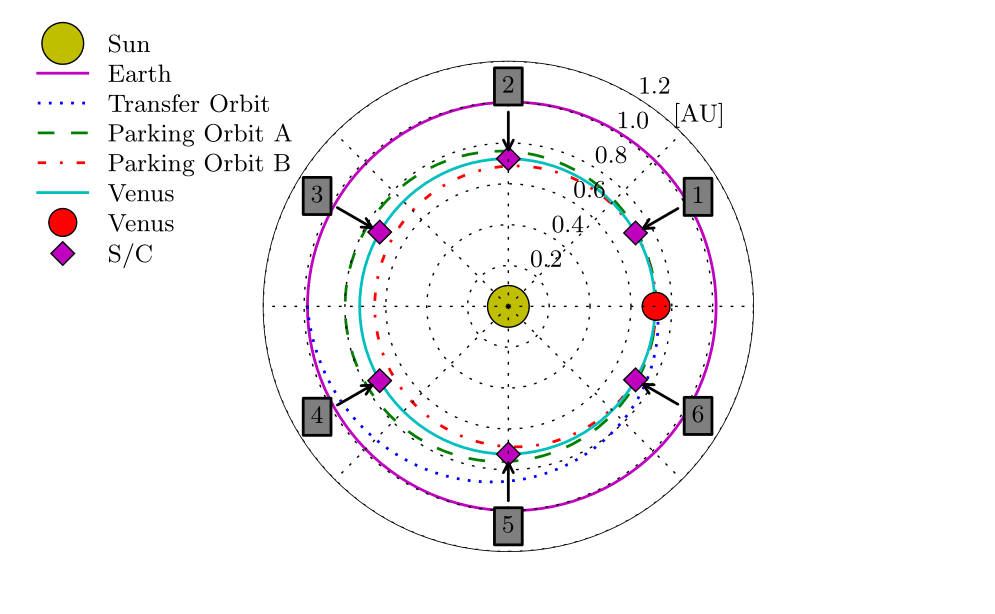}
\caption{The six spacecraft (S/C) are launched using two Ariane 5 rockets, two sets of three spacecraft. After launch the spacecraft are separated and travel towards Venus for a gravity assist maneuver (GAM). After the GAM the spacecraft will be parked into two orbits and will from thereon reach their final orbit, a Venus orbit with 60 degrees separation between each spacecraft.}
\label{fig:orbit}
\end{figure}

The use of parking orbits is essential in order to insert the six spacecraft into the distinct positions around the Venus orbit separated by 60 degrees and using two launchers and GAMs. These parking orbits have been designed to minimize the required propellant mass at minimal cost to the time taken to insert all spacecraft into the desired orbits.\\
The first three spacecraft will be launched into a transfer orbit from Earth's to Venus's orbital radius (Figure~\ref{fig:orbit}). During the transfer period the spacecraft will be in safe mode. After six months they will perform the GAM inserting two spacecraft into parking orbit A and one into parking orbit B. Parking orbit A is an elliptical orbit with perihelion equal to Venus's orbital radius and aphelion greater such that the period is 13/12 of Venus's orbital period. Similarly, parking orbit B is also elliptical, however the aphelion is equal to Venus's orbit and the perihelion is inside Venus's orbit such that the period is 11/12 of Venus's orbital period. The two spacecraft in parking orbit A will complete three and five full orbits until they reach orbital positions five and four, respectively. Similarly for the spacecraft in parking orbit B, which will complete five full orbits to reach position three. The last three spacecraft will be launched 19 months after the first three spacecraft and will follow the same transfer orbit to the gravity assist. Two spacecraft will be inserted into the parking orbit B and one into parking orbit A to fill the remaining positions. Once each spacecraft is in its final destination relative to Venus it can begin the one month commissioning phase; upon completion, operation can begin. The first two spacecraft will be operational 32 months after the first launch date, where the mission can become partially operational. Full operations can begin within less than 47 months after the first launch date.

\subsection{Spacecraft Design}
\label{Spacecraft Design}

\begin{figure}[h!]
   \centering
\includegraphics[width=0.6\textwidth]{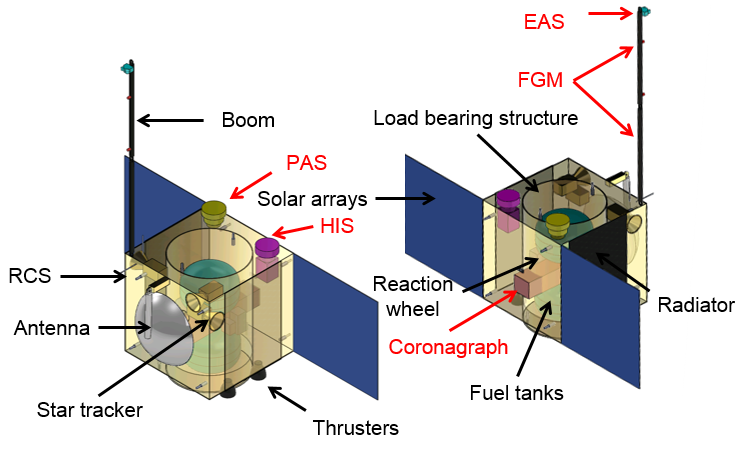}
\caption{Spacecraft design for the six identical spacecraft of the dimensions 1.5~m~$\times$~1.5~m~$\times$~1.7~m (yellow) excluding the solar panels (in blue, Section~\ref{Power}). The payload instruments are indicated in red text color (Section~\ref{Payload}) and the spacecraft subsystems in black text color.}
\label{fig:spacecraft_design}
\end{figure}

All six spacecraft will have the same design. Figure~\ref{fig:spacecraft_design} outlines the schematic view of one spacecraft that includes the sensors (indicated in red text color in the figure) that are described in Section~\ref{Payload} and some of the spacecraft subsystems (indicated in black text color), partly explained in Sections~\ref{Attitude Orbit Control System} to \ref{Thermal Control System}. The side of the spacecraft facing the Sun has an area of 1.5~$\times$~1.5~m$^2$, the dimensions towards space are 1.7~m.

\subsection{Payload}
\label{Payload}

The instruments employed for CARETAKER originate from other space missions with the intention to reduce mission costs and accelerate overall mission design. In-situ measurements are performed using the Solar Wind Analyzer instrumentation package from the \emph{Solar Orbiter} mission (Marsch et al. \cite{marsch2005}), together with flux gate magnetometers from \emph{Venus Express} (Titov et al. \cite{titov2006}), while remote measurements are performed using coronagraphs from the \emph{STEREO} mission (Kaiser et al. \cite{kaiser2008}).

\subsubsection{Coronagraph}
\label{Coronagraph}
Each spacecraft will include a coronagraph. Stereoscopic images of the Sun will be obtained by pairs of coronagraphs separated by 120 degrees. In order to obtain satisfactory stereoscopic images, each coronagraph will be switched off for the period the spacecraft orbits along the Sun-Earth line.
The coronagraphs used are externally occulted Lyot coronagraphs that observe from 2 solar radii to 15 solar radii. This choice of heliocentric distances provides the opportunity to observe the early stages of a CME.
The coronagraphs derive their heritage from the coronagraph COR2 on-board the SECCHI suit of instruments of the \emph{STEREO} mission (Howard et al. \cite{howard2008}). The external occultation shields the objective lens from direct sunlight, hence it enables a low stray light level and makes the observation in this range of heliocentric distances possible (Thernisien et al. \cite{thernisien2009}).
The technical features of the coronagraphs are presented in Table~\ref{tab:coronagraph}, where L$_0$ is the solar polarization brightness and DN is the measured response of the instrument in the passband of the observations.
The stereoscopic observations, combined with geometrical models (e.g. ice cream cone model, hollow croissant model) and reconstruction methods (forward modelling, inversion, and triangulation) allow to obtain the three-dimensional structure of CMEs and extract their propagation direction and velocity. The stereoscopic images obtained will assist to find the propagation direction and velocity of CMEs in the region 2-15 solar radii and hence to acquire an estimation of whether a CME will be Earth-directed or not.

\begin{table}[h!]
\centering
\caption{Coronagraph technical details.}
\begin{tabular}{ll}
\hline
Field Of View (FOV)&	11.4$^\circ$\\
Passband 	&450 nm -- 750 nm\\
Data rate 	&16.7 kbps\\
Compression factor	&10\\
Pixel size &	15 arcsec\\
Exposure time &	$<$4 sec\\
Image sequence cadence 	&5 min\\
Images per hour&	12\\
Photometric response (L$_0$/DN)	&1.3$\cdot$10$^{-12}$\\
\hline
\end{tabular}
\label{tab:coronagraph}
\end{table}

\subsubsection{Solar Wind Analyzer}
\label{Solar Wind Analyzer}
The Solar Wind Analyzer (SWA) consists of 3 sensors with a shared processing unit. The main objective of the SWA will be to characterize CME properties and to determine their magnetic field structure via comprehensive in-situ measurements of CMEs.
In order to meet all the measurement requirements, the SWA must be able to measure the three-dimensional velocity distribution functions of the major solar wind components: protons, electrons and heavy ions (ESA \cite{esa2011}).\\
The Electron Analyzer System (EAS) will make a high temporal resolution determination of the 3D electron velocity distributions and derive their moments (density, temperature, bulk velocity, heat flux).\\
The Proton Alpha Sensor (PAS) contains a top hat electrostatic analyzer that measures the 3D velocity distribution of protons and alpha particles in the energy range of 0.2 keV/q to 20 keV/q with a relative accuracy of 8\% in energy and an angular resolution of less than 2 degrees. Unlike the EAS, the PAS consists of only one device and has therefore a narrower field of view.\\
The Heavy Ion Sensor (HIS) contains an electrostatic analyzer module and a time of flight detector. It determines mass, charge, energy distribution, and direction of incidence of heavy ions up to 56 amu/q.\\
Table~\ref{tab:swa} lists the properties of the three sensors and their location onboard the spacecraft can be seen in Figure~\ref{fig:spacecraft_design}.

\begin{table}[h!]
\centering
\caption{Properties of the three sensors (Electron Analyzer System (EAS), Proton Alpha Sensor (PAS), and Heavy Ion Sensor (HIS)) that constitute the Solar Wind Analyzer (SWA).}
\begin{tabular}{cccc}
\hline
						&	EAS	&	PAS	&	HIS	\\
						\hline
						\hline
Field of View 	&	4$\pi$	& -24$^\circ$ -- 42$^\circ$ (Az)	&	 -30$^\circ$ -- 66$^\circ$ (Az)	\\
	 			&			&  -22.5$^\circ$ -- 22.5$^\circ$ (El)	&	  -17$^\circ$ -- 22.5$^\circ$ (El)	\\[0.2cm]
Particle Species		& Electrons	& H+, He++	& $^3$He -- Fe\\[0.2cm]
Energy Range			& 1 eV -- 5 keV	& 0.2 keV/q -- 20 keV/q	&	0.5 keV/q -- 100 keV/q (Az)\\
						& 				&						&	0.5 keV/q -- 16 keV/q (El)\\[0.2cm]
Energy Resolution 		& 12\%	&	8\%	&	6\%\\[0.2cm]
Measurement Parameters 	& flux, velocity,	&	3D velocity distribution	&	energy, charge,\\[0.2cm]
						& energy distribution 3D	&		&	mass, direction 3D\\[0.2cm]
Angular Resolution 		& 10$^\circ$	& 2$^\circ$	& 6$^\circ$\\[0.2cm]
Cadence 				& 4 sec	& 4 sec	& 30 sec\\
\hline
\end{tabular}
\label{tab:swa}
\end{table}

\subsubsection{Fluxgate Magnetometer}
\label{Fluxgate Magnetometer}
The Fluxgate Magnetometer (FGM) will perform in-situ measurements of the magnetic field vector. The instrument uses two triaxial fluxgate sensors which allow separation of stray field effects of the spacecraft from the ambient magnetic field. One of the sensors is mounted on a boom with a length of 3 m, while the other sensor is directly attached to the spacecraft.
The sampling rate is $1~\mathrm{Hz}$ for normal operation. Values are averaged to one minute values.
The instrument design is based on the Fluxgate Magnetometer (MAG, Zhang et al. \cite{zhang2007}) on \emph{Venus Express}, except for increased boom length. Similar instruments have also been flown on \emph{Rosetta Lander} (Biele and Ulamec, \cite{biele2007}) and the Mir Space station. This instrument is capable of measuring in interplanetary space as well as inside the magnetic field of Venus (Balogh \cite{balogh2010}). The range of the outer sensor is $\pm 262~\mathrm{nT}$ with an accuracy of $8~\mathrm{pT}$ and the on-board sensor with a range of $\pm 524~\mathrm{nT}$, also with an accuracy of $8~\mathrm{pT}$. This theoretically available accuracy exceeds the required resolution of $100~\mathrm{pT}$ (Section~\ref{Mission Statement and Requirements}) and is therefore not needed at such a high level, which would increase mission costs unnecessarily.

\subsection{Attitude Orbit Control System (AOCS)}
\label{Attitude Orbit Control System}
All of the spacecraft will be 3-axis stabilized. The alternative (spin stabilized) maintains a fixed orientation with respect to the stars and is therefore not suitable for this mission. Each spacecraft must face the Sun at all times and subsequently must complete one full rotation in sync with its orbital period. In addition, to transmit to Earth consistently, the antenna must be steerable for pointing, which is only feasible with 3-axis stabilized configuration. The control system for stabilization will consist of momentum wheels\footnote{http://www.rockwellcollins.com/sitecore/content/Data/Products/Space\_Components/Satellite\_\\
Stabilization\_Wheels/RSI\_12\_Momentum\_and\_Reaction\_Wheels.aspx (accessed July 2013).
} in conjunction with thrusters\footnote{http://cs.astrium.eads.net/sp/spacecraft-propulsion/bipropellant-thrusters/4n-thruster.html (accessed July 2013)}$^,$\footnote{http://cs.astrium.eads.net/sp/spacecraft-propulsion/bipropellant-thrusters/22n-thruster.html (accessed July 2013)} in the reaction control system (RCS) for de-saturation. The various components are detailed in the following sections.

\subsubsection{AOCS Components}
\label{AOCS}
In order to optimize stability while accounting for redundancy the reaction wheels are mounted in a tetrahedral configuration (Wagner et al. \cite{wagner2012}); attitude control can be achieved with four wheels operating simultaneously (the nominal operational scenario) or any combination of three wheels (if one wheel were to fail).

\noindent
Three star trackers\footnote{http://www.selex-es.com/domains/space/attitude-control-sensors (accessed Feb.~2014)} (STRs), one in cold redundancy, determine the orientation relative to the stars. Two inertial measurement units (IMU) are mounted on a common platform, providing a finer control of orientation. These IMUs consist of three Fibre Optic Gyros for rate measurement (Hablani \cite{hablani1994}). One is required for normal operation and the second is for redundancy. Sun sensors determine the spacecraft orientation relative to the Sun and allow for attitude control in safe mode, in which thrusters are used to keep solar cells aimed at the Sun and avoid damaging any vulnerable imaging instruments. There are a total of two Sun sensors$^4$ (one redundant) and four reaction wheels (one redundant). \\
During science operations, at least two STRs will be used in combination. In the event of major system anomaly in the spacecraft and consequent loss of attitude control, dedicated shutters will protect the STR optical paths to prevent damage due to accidental Sun pointing. Orientation algorithms will be processed on the spacecraft's main CPU.\\

\subsubsection{Reaction Control System (RCS)}
\label{Reaction Control System}

\noindent
The spacecraft are required to be separated by 60 degrees in their  orbit around the Sun. As discussed in Section 3.1, the period of the transfer orbits is crucial to allow for efficient insertion into the desired orbit with the correct phase.

\noindent
There will be 24 RCS thrusters arranged in groups of three at the corners of the main cubic structure of each spacecraft. The thrusters of one block form a triad, thereby producing thrust in the three perpendicular directions. Four blocks are sufficient to allow the spacecraft to rotate and translate in any direction. Eight sets of thrusters thus grant full, built-in redundancy. The RCS momentum dumping will require only short, sporadic pulses. The total burn time estimated for the RCS thrusters is calculated at 10.2 hours with cycles per thruster not exceeding 10\% of the rated capacity of the thrusters.

\noindent
Table~\ref{tab:orbitpertub} shows the propellant mass required to maintain the orbit of each individual spacecraft. Perturbation calculations were completed with respect to a full solar cycle ($11\times 365$ days).

\begin{table}[h!]
\centering
\caption{Breakdown of the mass of propellant required for station keeping for each of the 6 spacecraft. The spacecraft are listed in order of heliocentric angular separation from Venus. The associated propellant mass required for station keeping, over the lifetime of the mission, is shown per spacecraft.}
\begin{tabular}{ccc}
\hline
Spacecraft number & Angular separation from Venus & Propellant mass for station keeping \\
 & $\left[ \mathrm{^\circ deg}\right]$ & $\left[ \mathrm{kg}\right]$ \\
\hline \hline
1 & 30 & 16.6 \\
2 & 90 & 19.9 \\
3 & 150 & 16.8 \\
4 & 210 & 14.0 \\
5 & 270 & 16.8 \\
6 & 330 & 18.0 \\
\hline
\end{tabular}
\label{tab:orbitpertub}
\end{table}

\subsubsection{Propulsion System}
\label{Propulsion System}
The propulsion systems considered for the CARETAKER mission include chemical, electrical and hybrid options. Electrical propulsion has the potential to save more than 100 kg per spacecraft with respect to chemical propulsion. However, electric propulsion has low thrust capabilities and therefore longer transfer times as well as a high power consumption. It consequently requires extra solar arrays and the resulting high thermal output has furthermore implications on the thermal system. Additionally, being a relatively new technology it harbors an increased risk of failure. \\
A third option is a hybrid solution. This would involve employing chemical thrusters to provide the major burns, then utilizing low thrust electric thrusters to maneuver into the final position. The RCS uses chemical thrusters therefore, compared to a purely electrical system, the increase in mass of additional equipment would be minimal. Calculating the optimal chemical and electric propulsion hybrid along with the subsequent orbital insertion procedure is beyond the scope of this paper but there is potential to reduce the overall insertion time and mass.\\
After considering the above, it was concluded that propulsion system will be chemical. The total mass of propellant required for the insertion of each spacecraft into its desired orbit is 310 kg. Each spacecraft will have 22 N thrusters which allows them to perform the necessary burns for the orbital maneuvers in about two hours. With an accumulated burn life of 70 hours and 1,000,000 cycles, the chosen thrusters are well within the operational limits.

\subsubsection{Propellant}
\label{Propellant}
RCS and main engine thrusters will be supplied from a common bi-propellant tank set. Propellant requirements have an optimal mix ratio of 1.4 and the total mass requirement amounts to 306.2~kg.\\%
Two bladder tanks were selected, allowing for sufficient expansion. The specific tanks selected were Astriums 198L bladder tanks\footnote{http://cs.astrium.eads.net/sp/spacecraft-propulsion/propellant-tanks/198-litre-bipropellant-tank.html (accessed Feb. 2014)}.%

\subsection{Power}
\label{Power}

The electrical power sub-system provides, stores, distributes and controls the spacecraft energy. The primary power sources considered are the solar panels; batteries are the secondary power source.

\subsubsection{Primary Power Sources -- Solar Panels}

The solar panels convert  the radiation from the Sun into electrical power to maintain full operation of the sub-systems. Included in the trade-off analysis for the solar panels are the spacecraft orbit and the power requirements of the sub-systems.\\
Hence, the best option in terms of efficiency and dimensions are gallium-arsenide semiconductor cells which double the efficiency of a silicon solar cell used in the past. 28\% Triple Junction GaAs Solar Cell of Type: TJ Solar Cell 3G28C\footnote{http://www.azurspace.com/images/products/HNR\_0002490-00-03.pdf} were considered the best solution for the spacecraft. With respect to the total amount of power needed to operate each spacecraft, the required area of the solar panels is approximately 3~m$^2$.

\subsubsection{Secondary Power Sources -- Batteries}
Batteries are used as a secondary power system, operating to store energy. Taking into account the available space-rated batteries and the mission requirements, the energy is stored in nickel-hydrogen batteries. With space-based communications requiring large amounts of power and high reliability, the $\mathrm{NiH}_2$ batteries represent a desirable power supply with a long cycle life (Dermott et al. \cite{dermott1996}).
During the launch, the CPU and the communication requires up to 30~W, thus leading to a discharge of 6\% after 2 hours. In safe mode, the batteries will still be charged at 52\% after 24 hours.\\
The secondary power supply will contain 12 cells of $\mathrm{NiH}_2$, to provide approximately 100 Wh per cell.

\subsection{Thermal Control System}
\label{Thermal Control System}
The thermal control system has been designed to ensure that the spacecraft instrument and component temperatures are always within the operational range of 298 K $\pm$ 15 K during the orbital periods and operational lifetime. A total of 2.2 m$^2$ of the deep-space facing sides will be coloured matt-black to radiate excess heat. The remainder of the spacecraft surface will be covered in multi-layer insulation (MLI). Fluid pipes will be employed to transport excess heat from the instruments and the Sun facing side to the radiator panels. The spacecraft will be actively heated during the cold phases where the instrument power is not heating the spacecraft (safe mode and transfer from Earth orbit to Venus orbit). A maximum of 260 W will be required to maintain the spacecraft within the acceptable temperature range. This power will be drawn from either the batteries or the solar panels and will be fed to resistive heaters.

\subsection{Radiation Environment and Shielding}
\label{Shielding}
\begin{figure}[h!]
   \centering
\includegraphics[width=0.51\textwidth]{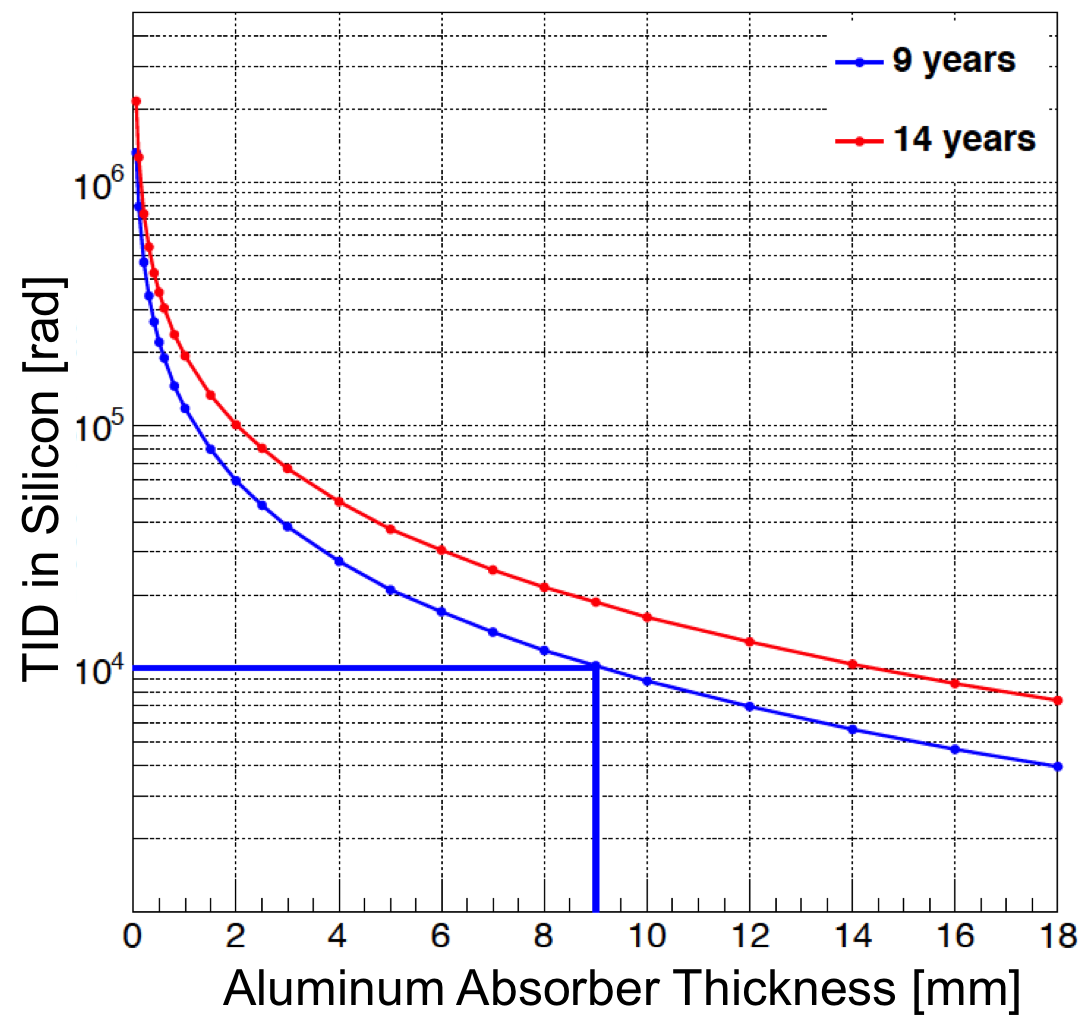}
\caption{Total ionizing dose in silicon vs. aluminum absorber thickness inside solid aluminum spheres calculated with SPENVIS. The blue lines indicate the required limit of 10~krad corresponding to a shielding thickness of 9~mm aluminum.}
\label{fig:tid}
\end{figure}

During the mission all spacecraft encounter the complex and harsh radiation environment in space composed of galactic cosmic rays (GCR) and solar energetic particles. All of the spacecraft will be briefly exposed to significantly increased particle fluxes as they cross Earth's radiation belts during the orbital transfer. Electronic equipment is particularly vulnerable to radiation exposure, therefore the spacecraft will be equipped with radiation hard electronics where possible. Aluminum shielding will be implemented to reduce the risk of spacecraft failure. In order to account for the increased radiation exposure in the space environment, a preliminary shielding study is performed employing the 'radiation sources and effects' package in the online tool SPENVIS\footnote{http://www.spenvis.oma.be, accessed Sept. 2013} (Space Environment Information System). Solar protons are dominating, hence only solar energetic particles are considered in this analysis. As an upper limit for the total ionizing dose (TID) the time period between the launch date and the end of the five-year mission operation (9 years in total) is considered at a distance from the Sun of 0.7 AU. Figure~\ref{fig:tid} shows the dose as a function of aluminum shielding thickness for nine years and for the extended mission period of 14 years in total using the SHIELDOSE-2 model within the 'long-term radiation doses' package. All electronic components shall tolerate a TID of 20 krad, therefore the dose in silicon is calculated. The shielding will be designed to yield a TID exposure of 10 krad, providing a factor 2 margin (Wertz and Larson \cite{wertz1999}). This results in an estimated thickness of 9 mm for the nominal mission lifetime. Under the assumption that the spacecraft itself provides a minimum shielding of 3 mm aluminum, all critical components are shielded with an individual aluminum envelope of 6 mm thickness. The estimated shielding mass calculates to 43 kg for each spacecraft (Table~\ref{tab:mass_budget}).

\noindent
Studies of the total non-ionizing dose (TNID) are of special importance for the degradation and therefore for the performance of the solar panels. The EQFLUX package within SPENVIS has been used to calculate the damage equivalent fluences. The thereby obtained results for the mission duration is compared to the respective solar panel properties (Section~\ref{Power}). The solar panels consequently show an estimated loss of efficiency of less than 2\% until the end of the nominal operational time.\\

\section{Communication Segment}
\label{Communication Segment}

\subsection{Communications Scenarios and Subsystem}
\label{Communications Scenarios and Subsystem}

\begin{figure}[H]
   \centering
\includegraphics[width=0.7\textwidth]{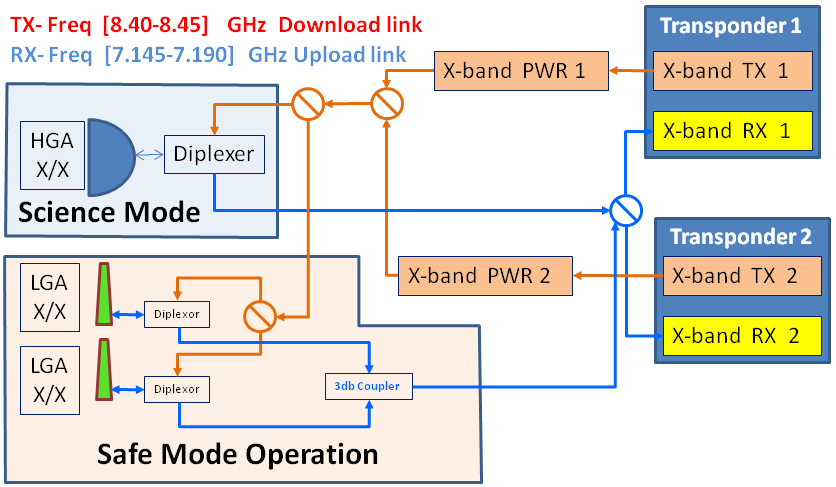}
\caption{Schematic overview of the onboard communication systems. Each spacecraft has a full duplex communication system and consists of two operational modes, a safe mode using two low gain antennas (LGA) and a science mode using one high gain antenna (HGA).}
\label{fig:comm_opmodes}
\end{figure}

There are two communication scenarios, each corresponding to a different mission phase (see Figure~\ref{fig:comm_opmodes}).
\begin{enumerate}
	\item Safe mode: launch and early orbit phase (LEOP) and cruise phases\\ The downlink data consists of housekeeping information and science data. Communication via two wide beam-width low gain antennas (LGAs) using X-band is considered as baseline for 7.145-7.190~GHz and 8.40-8.45~GHz for up and downlink respectively (Fortescue et al. \cite{fortescue2011}). The LGA provides omni-directional coverage and telemetry (TM) up to a distance of 0.8 AU from 15 m antennas on Earth.
	\item Science mode: nominal and extended phases\\ A simple pointing mechanism will be used (Noschese et al. \cite{noschese2011}) on the deployable high gain antenna (HGA), with 1 m diameter and 1.75 degrees beam-width.
\end{enumerate}

\noindent
Additionally, the communication subsystems will include a hot redundant set of X/X Band transponders and 60 W traveling wave tube amplifier (TWTA) and a RF distribution unit (RFDU) with diplexers, 3dB couplers and wave guides to provide the nominal communication with Earth during the duplex spacecraft-Earth Station connection phases of the mission (Figure~\ref{fig:comm_earth_spacecraft}).

\subsection{Earth Stations}
\label{Earth Stations}

\begin{figure}[H]
   \centering
\includegraphics[width=0.9\textwidth]{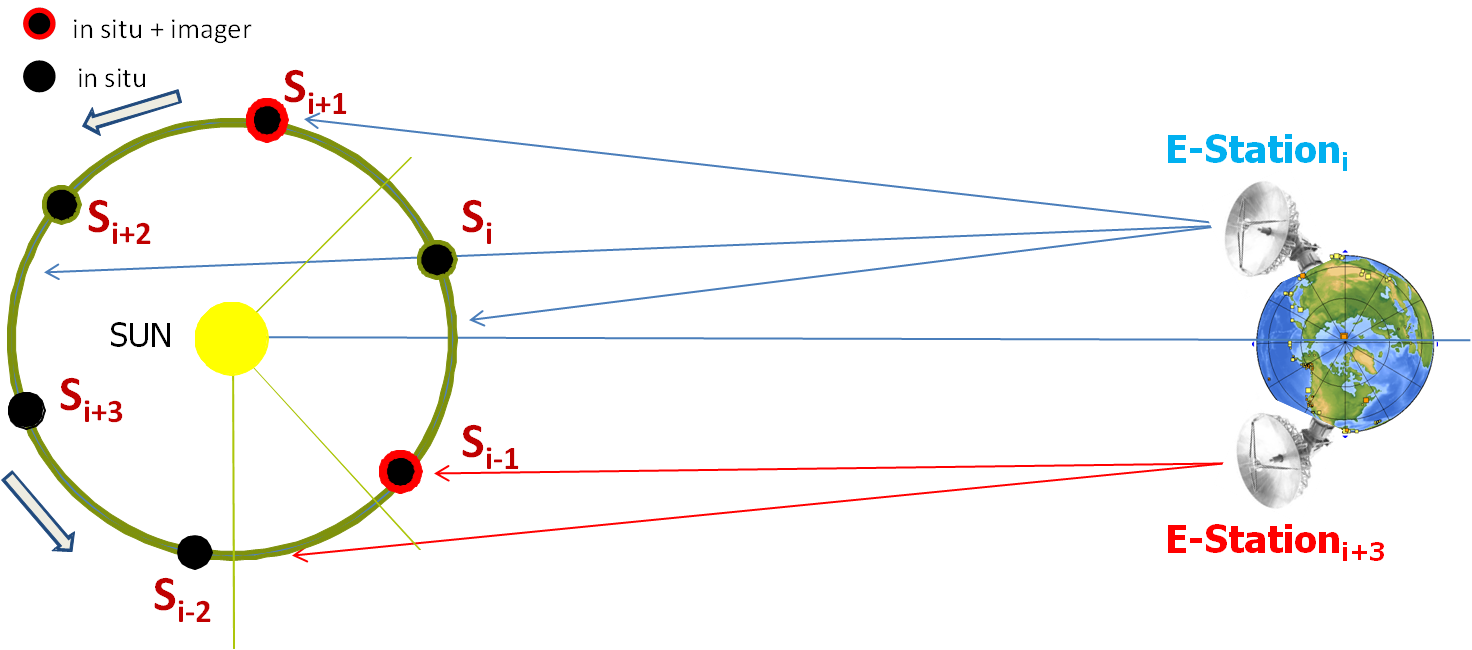}
\caption{Illustration of Earth--spacecraft communication and instrument activity. The deep space network at Earth consists of six antennas of which two (E-Station$_{i}$ and E-Station$_{i+3}$) are operating simultaneously as displayed in the figure. Each Earth station is intermittently in contact with only three spacecraft, all spacecraft perform in-situ measurements whereas only two spacecraft also perform remote measurements (i.e. in-situ + imager), depending on their location.}
\label{fig:comm_earth_spacecraft}
\end{figure}

\begin{figure}[H]
   \centering
\includegraphics[trim=4cm 3cm 2cm 1cm, clip=true, width=1.\textwidth]{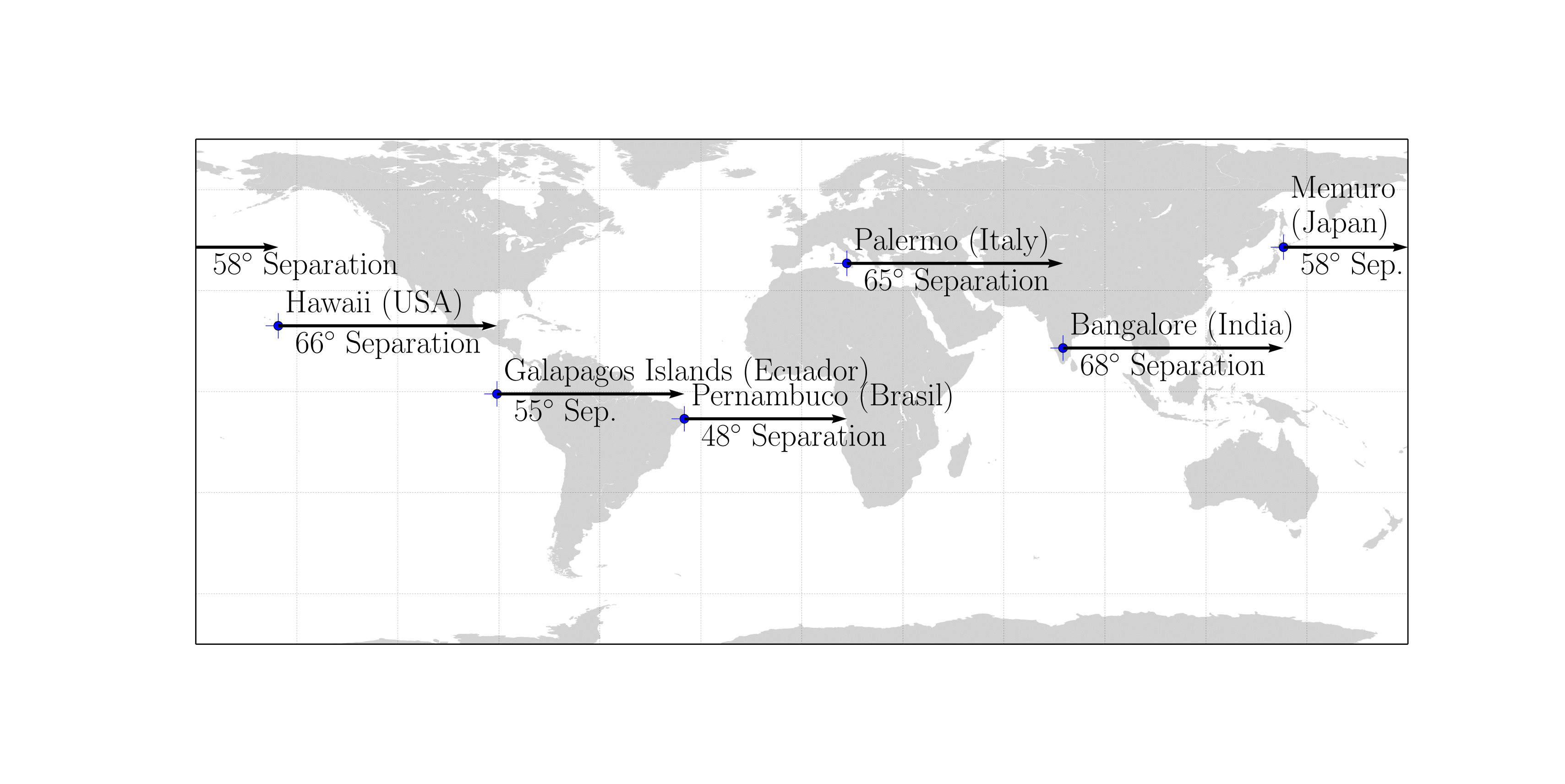}
  \centering
\caption{This overview shows the longitudinal locations of the six 15 m diameter dishes that are part of the deep space network; these locations ensure simultaneous operation of two dishes for contacting the spacecraft.}
\label{fig:earth_stations}
\end{figure}

Science and housekeeping data is received by two antennas in parallel at the same time from both sides of the Earth, (Figure~\ref{fig:comm_earth_spacecraft}). The uplink and downlink will use X-band. It is planned to distribute the stations equally over the Earth to ensure parallel download from the spacecraft. In order to ascertain a continuous data stream from the spacecraft, six dedicated 15 m dishes will be used to download the science and housekeeping data.\\
Due to high priority data downloading and time consuming antenna pointing tasks, there is no possibility to reuse the existing ESTRACK Deep Space Network (DSN) \footnote{http://www.esa.int/Our\_Activities/Operations/Estrack\_tracking\_stations (accessed Feb.~2014)}, NASA's DSN \footnote{http://deepspace.jpl.nasa.gov/ (accessed Feb.~2014)} or the Indian DSN
 \footnote{http://www.isro.org/GroundFacilities/trackingfacility.aspx (accessed Jun.~2014)}.
 Figure~\ref{fig:earth_stations} shows the proposed longitudinal locations of the Earth stations that are part of the DSN defined for this mission. \\

\section{Operation and Ground Segment}
\label{Operation and Ground Segment}
\begin{figure}[h!]
   \centering
\includegraphics[width=1.0\textwidth]{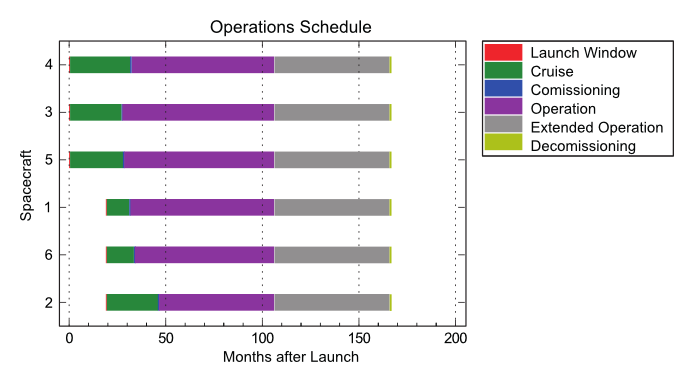}
\caption{Life cycle diagram for all spacecraft, showing that the six spacecraft are not launched simultaneously but in two sets of three, at different launch dates. The differences in cruise duration for individual spacecraft (e.g. Spacecraft 2 compared to Spacecraft 1) is required for the 60 degree spacecraft separation. The spacecraft numbering refers to the numbering introduced in Figure \ref{fig:orbit}.}
\label{fig:mission_phases}
\end{figure}

As discussed in Section~\ref{Orbit and Launch}, every 19 months there is a three week window to launch this mission. The mission timeline is depicted in Figure~\ref{fig:mission_phases}, showing the time separated launch at a nominal launch date of the spacecraft in two sets of three spacecraft, the individual cruise duration for each spacecraft, commissioning and nominal operation in the final orbits. Subsequently the optional extended mission timeline and decommissioning are displayed. The spacecraft numbering in the figure refers to the numbering in Figure \ref{fig:orbit}.

\begin{figure}[h!]
   \centering
\includegraphics[width=0.6\textwidth]{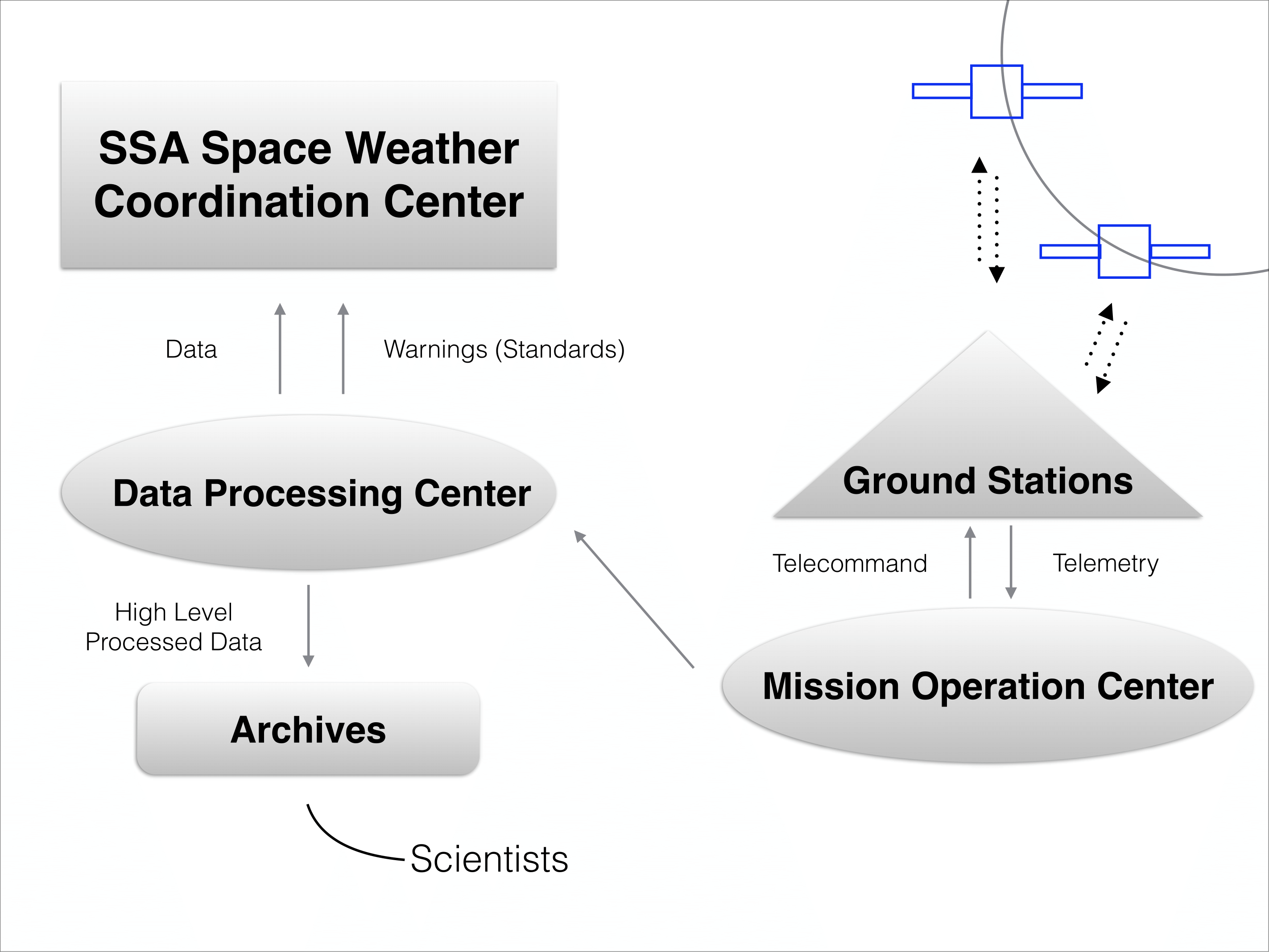}
\caption{Illustrating data handling at Earth. Raw data is obtained through six ground stations and is collected by a mission operation center that forwards the data to a data processing center. This center processes the data and provides the SSCC Space Weather Coordination Center with data according to their format, additionally, the data is also archived for the scientific community. It is eventually the SSCC that provides the public with a forecast.}
\label{fig:Ground_segment}
\end{figure}

\noindent
The mission operations concept shall minimize the costs both in the area of ground segment tools and facilities as well as in the sharing of manpower and expertise in the development and operations teams. \\
It is important to simultaneously approach the spacecraft system-level testing between the spacecraft manufacturer and the spacecraft operations team, maximizing the synergy between spacecraft manufacturer and operators in the preparation of operational documentation, spacecraft user manual, operations database etc.\\
The ground segment will rely on six ground stations all around the world with two of them continuously communicating with the spacecraft (Section~\ref{Earth Stations}).  The ground stations must be built for the mission since the capacity of any existing DSN is insufficient for a continuous link. Figure~\ref{fig:earth_stations} shows the proposed locations of such a network. The objective is to build stations able to command the spacecraft and to receive data. For a given time two of them will be absolutely dedicated to communication with CARETAKER. However there will always be four stations available for other purposes. The ground segment organization including the further data transfer and processing is presented in Figure~\ref{fig:Ground_segment}.

\newpage

\section{Data Processing}
\subsection{Mission Operation Center}
\label{Mission Operation Center}
\begin{figure}[h]
   \centering
\includegraphics[width=0.8\textwidth]{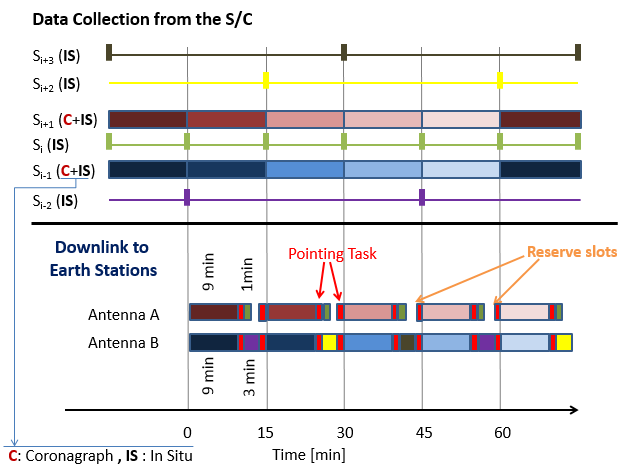}
\caption{Data collection from the spacecraft:
Colors stand for the data collected by each spacecraft communicating with the ground segment. Data from four spacecraft is downloaded every 15 minutes with two antennas on Earth. For each such interval the downlink time from spacecraft sending imaging and in-situ (C+IS) data takes nine minutes. The antennas then switch to the next spacecraft through a one minute pointing procedure and download only the in-situ (IS) measurements from the other spacecraft.}
\label{fig:data}
\end{figure}

The CARETAKER mission operation center (MOC) will be in charge of all telecommand and telemetry operations of the mission. Communication with the six spacecraft will be provided by two ground stations at a time through the CARETAKER Network 24/7. The detailed communication schedule is given in Figure~\ref{fig:data}. The two antennas that communicate with the spacecraft at a given time download data from four spacecraft every 15 minutes. For each such interval, each antenna will download data from two spacecraft, one providing imaging and in-situ data (C+IS), the second only sending in-situ (IS) data. While data from the two spacecraft sending imaging (S$_{i+1}$ and S$_{i-1}$ in the figure) data and the spacecraft within the Sun--Earth line (S$_{i}$) is downloaded in every interval, the other three spacecraft are not necessarily addressed every 15 minutes. \\
The operation center will perform 0-level data processing as well as backing up and storing the previous 30 days of the processed data. Data will be transmitted continuously to the data processing center. Data processing for in-situ measurements is likely to take less than one minute so that the requirements of the SSCC are fulfilled.

\subsection{Data Processing Center}
\label{Data Processing Center}
The data processing center shall perform different tasks:\\
\begin{enumerate}
	\item Calibration, validation and processing of the data for the early phase of the mission.
	\item Provide algorithms, methods and models to a warning unit which shall be able to give standards compatible data to the SSCC. The time between taking the measurement and delivering the data products should meet those standards (see Section~\ref{Data Products}).
	\item Perform higher level processing on the data and make it available on the Internet.
	\item Archive all data for further and long term investigation (over a semi-solar cycle as it is possible with the minimum life time of the mission).
\end{enumerate}

\subsection{Data Products}
\label{Data Products}
Given the extraordinary position of the spacecraft constellation, the processed data is to become a new reference for space weather event warning as well as premium scientific content. For the in-situ measurement, a fifteen minutes range between measurement and data product will be ensured as required by the SSCC. For 3D-modelling of coronagraphic data and CMEs, an extended time line is proposed: those data should be available within a maximum of sixty minutes after remote measurement.
Further processing will be performed later to make the data fully exploitable by the scientific community in the long range. The whole content will be available on the internet.

\section{Budgets}
\label{Budgets}
Budgets concerning mass and power of the spacecraft are analyzed considering the mission profile and the lifetime of the mission. The launcher fairing size and mass constraints were taken into account and all spacecraft were investigated individually. Margins for each subsystem are applied following the ESA Margin Philosophy for Science Assessment Studies (SCI-PA/2007/022, \cite{margin2007}). According to this, off-the-shelf components that are implemented without any changes are considered with a margin of 5\%, a 10\% margin is added for off-the-shelf items with minor modifications and a 20\% margin is used in case of major modifications, new designs and new developments. Additional 20\% system margin are added in the end to the complete system. The resulting mass and power budgets are summarized in Table~\ref{tab:mass_budget} and Table~\ref{tab:power_budget}, respectively.

\noindent
The propellant in the mass budget includes additional fuel for AOCS/RCS and safe mode recovery as well as a margin of 20\%. As this budget refers to a single spacecraft, the total launch mass within one Ariane~5 rocket adds to 2812~kg, including three spacecraft plus launch adapters.

\label{Mass Budget}

\begin{table}[h!]
\centering
\caption{Mass budget for a single spacecraft including the relevant margins.}
\begin{tabular}{lccc}
Subsystem	&	Mass	&	Margin	&	Mass total	\\
	&	[kg]	&		&	[kg]	\\
\hline
\hline
Power	&	43.00	&	2.75	&	45.75	\\
Payload	&	31.00	&	2.96	&	33.96	\\
Communications	&	28.20	&	5.22	&	33.42	\\
Onboard Data Handling / Avionics	&	15.00	&	0.75	&	15.75	\\
AOCS	&	92.86	&	2.45	&	95.31	\\
Thermal Control	&	24.00	&	1.40	&	35.20	\\
Additional Shielding	&	42.67	&	0.00	&	42.67	\\
Chemical Propulsion System (dry mass)	&	37.60	&	3.63	&	41.23	\\
Harness (5\%)	&	15.72	&	0.00	&	15.72	\\
Structure (20\% of dry mass)	&	66.01	&	0.00	&	66.01	\\
\hline
TOTAL (dry, without system margin)	&	396.05	&	22.18	&	418.23	\\
System Margin (20\%)	&	79.21	&		&		\\
TOTAL (dry, with margin)	&		&		&	497.44	\\
Propellant	&		&		&		306.2 \\	
\hline
TOTAL (wet mass)	&		&		&	803.64 \\		
\hline
\end{tabular}
\label{tab:mass_budget}
\end{table}

\label{Power Budget}

\begin{table}[h!]
\centering
\caption{Power budget for a single spacecraft including the relevant margins.}
\begin{tabular}{lccc}
Subsystem	&	Power Consumption	&	Margin	&	Power Consumption total	\\
	&	[W]	&		&	[W]	\\
\hline
\hline
Power	&	48.0	&	0.0	&	48.0	\\
Payload	&	18.5	&	1.8	&	20.3	\\
Communications	&	165.0	&	16.3	&	181.3	\\
Onboard Data Handling / Avionics	&	17.0	&	0.85	&	17.85	\\
AOCS	&	198.6	&	99.3	&	297.9	\\
Thermal Control	&	260	&	13	&	273	\\
Chemical Propulsion System	&	5.0	&	1.0	&	6.0	\\
\hline
TOTAL (without margin)	&	712.1	&	132.2	&	796.3	\\
System Margin (20\%)	&	142.4	&		&		\\
\hline
TOTAL (with margin)	&		&		&	986.67	\\
\hline
\end{tabular}
\label{tab:power_budget}
\end{table}

\newpage

\section{Risk Analysis}
\label{Risk Analysis}
As per ECSS (European Cooperation on Space Standardization\footnote{http://www.ecss.nl/}) standard operating procedures, outlined in ECSS-M-ST-80 (\cite{ecss2008}), a risk matrix was developed to classify potential risks for the mission. There are two main sub-classifications under which each risk will be listed; their likelihood (A [low] to E [high]) and their severity (1 [low] to 5 [high]). Risk is calculated as a combination of severity and likelihood; the result ranges from A1 [very low] to E5 [very high]. The high and very high risks are of the utmost concern and if not possible to reduce their classification they may represent reasons for mission postponement until such time as the level can be reduced. In extreme cases it can also lead to overall mission cancellation. Due to the complexity of this multi-spacecraft mission Table~\ref{tab:risks} shows the risk analysis of a subset of cases and their respective countermeasures.

\noindent
The dependence of the primary mission objective on the performance of all six spacecraft produces the most severe risk: the failure of only one spacecraft compromises the mission, as due to the incomplete coverage of the in-situ measurements, a CME heading to Earth might be missed. This problem could be mitigated by employing a larger number of spacecraft, which would increase costs significantly.

\begin{table}[t]
\centering
\caption{Risk examples for the mission classified following ECSS-M-ST-80 (\cite{ecss2008}) for their their likelihood (A [low] to E [high]) and their severity (1 [low] to 5 [high]).}
\begin{tabular}{p{4.5cm}ccp{7cm}}
Possible Scenario	&	\multicolumn{2}{c}{Risk Index}	&	Proposed action	\\
\hline
\hline
Launch fails	&	A5	&	low	&	No action - mission failed	\\[0.2cm]
Failure of one spacecraft	&	B5	&	medium	&	Continue mission with remaining spacecraft, focus on scientific research\\[0.2cm]
Launch misses time window	&	C2	&	low	&	Wait 19 months for the next time window	\\[0.2cm]
Miss the trajectory for GAM	&	B5	&	medium	&	mission failed\\[0.2cm]
Collisions of spacecraft during deployment	&	B4	&	low	&	Communicate to spacecraft and try to recover possible orbit problems with propulsion system\\[0.2cm]
Not enough data rate during safe mode to ensure communication between spacecraft and ground station	&	B3	&	low	&	Spacecraft are on a circular orbit pointing to the Sun. Wait until spacecraft is closer to the ground station, where higher data rate is possible. Two small emergency antennas (70 m dish ground station)	\\[0.2cm]
Damage of sensitive optics of a coronagraph during space flight	&	B3	&	low	&	Analyze the influence of the damage and try to reduce it using image processing at the science ground station	\\[0.2cm]
Radiation damage of the measurement system caused by GCR background and SEPs	&	D2	&	medium	&	Expected damage, which limits the lifetime of the measurement system. Possible to move spacecraft slightly in order to protect them.	\\[0.2cm]
If the spacecraft enters safe mode in a 40 degrees range on the backside of the Sun, connection will not be possible in this area	&	B3	&	low	&	Ask communication specialist. Communicate with spacecraft, when it leaves the 40 degrees area behind the Sun and recover it	\\[0.2cm]
Blackout of measurements device	&	A3	&	low	&	Ask payload specialist. Check the whole spacecraft housekeeping system, special communication time to the spacecraft	\\[0.2cm]
Explosion of some propulsion tank after deployment	&	A4	&	low	&	Check damage, continue mission with remaining spacecraft	\\
\hline
\end{tabular}
\label{tab:risks}
\end{table}

\section{Cost Analysis}
\label{Cost Analysis}
In comparison to typical ESA science missions, the CARETAKER mission is expensive. This is due to the mission objective of creating a forecast system which is not purely scientific, but shall rather serve as a warning system in order to prevent possibly catastrophic consequences on Earth, of which some have been discussed before. Such a warning system however requires that events with a possible impact can be observed reliably. This can be only achieved with significantly more resources than are available for purely scientific missions. A smart resource management is required to reduce the costs of the mission in a reliable frame. The total costs are estimated to lie between 1.2 billion Euro and 1.4 billion Euro. About 50\% of the budget will be spent on design and construction of spacecraft platform and payload. This would be much higher if the spacecraft were not designed to be identical.  Approximately 30\% of the costs will be taken to provide the infrastructure of the ground segment and the operation service. This part is relatively large, since the mission requires its own deep space network. The main uncertainties within the cost approximation are caused by the ground segment. There is no experience for multi-spacecraft missions in deep space with a continuous communication coverage. The architecture of the spacecraft allows to pack the complete set of spacecraft into two Ariane~5 launchers. The launch takes the smallest portion of the cost with 20\%.\\
In comparison, the JUICE mission (Grasset et al. \cite{grasset2013}), costs ESA 870 million Euros and Rosetta\footnote{http://sci.esa.int/rosetta/} is estimated to cost 1.4 billion Euros (according 2014 economic conditions\footnote{http://www.esa.int/Our\_Activities/Space\_Science/Rosetta/Frequently\_asked\_questions, What is the total mission cost?}). CARETAKER can only be funded by an international collaboration of space agencies and economic companies.

\section{Conclusion}
The presented mission concept consists of six identical spacecraft, located in a heliocentric orbit at 0.72 AU at a separation angle of 60 degrees. It is aimed to become fully operational (i.e. with all spacecraft in place) by 2030, while partial operation with less spacecraft can already begin three years prior to this. The nominal mission life time with all spacecraft is 5 years with a possible extension of additional 5 years, covering an entire solar cycle.\\
The mission's primary objective is to provide a near real-time information-service for space weather, based on the physical properties of solar CMEs. The information will consist of both in-situ and remote measurements. The mission is optimized to monitor CMEs with an angular extent of 60 degrees and larger, since these have the most potential to cause geomagnetic storms upon impact with Earth. The most important information will contain the propagation velocity and direction of a CME, as well as the spatial resolution of the magnetic field orientation and strength. This information helps in predicting the time of impact at Earth, and the potential severity of a geomagnetic storm if the CME proves to be geoeffective.\\
The proposed instruments to be employed on each spacecraft originate from other space missions with the intention to reduce mission costs and to accelerate overall the mission design. The spacecraft will perform in-situ measurements using the Solar Wind Analyzer instrumentation package from the \emph{Solar Orbiter} mission, together with flux gate magnetometers from \emph{Venus Express}, while performing remote measurements with a coronagraph from \emph{STEREO}.\\
The mission's secondary objective is to provide data helping to improve CME models. As for the choronagraph the present limitation for the 3D-reconstruction for the CMEs is given by the two \emph{STEREO} spacecraft and \emph{SOHO} which are not always positioned in a useful way to provide the necessary view angles (Mierla et al. \cite{mierla2010}). For the in-situ measurements models exist for determining the flux rope geometry, which in the case of one spacecraft this represents a 1D local cut through a global 3D structure. Having more than one spacecraft sampling the CME could help to constrain its global structure (M{\"o}stl et al. \cite{moestl2012}).\\
Communication is realized via a dedicated deep space network consisting of six ground stations equipped with 15 m dish antennas, with continuous operation of two ground stations simultaneously, each of them alternating contact between three spacecraft. Received data will be processed by a data processing center that forwards the processed data to the Space Weather Coordination Center who eventually informs the general public through a space weather forecast. The data processing center will additionally archive the data for the scientific community, for updating scientific models.

\noindent
The obtained information will result in major advances in space weather forecasting time and the scientific modelling of space weather. The data provided can be combined with data from other missions running at the same time, which has the potential to further expand the solar physics understanding. Furthermore, the proposed mission will not only monitor the solar wind in the Sun -- Earth direction, but it also offers the potential for 360 degrees warnings, providing information to a warning system that protects astronauts in future manned missions (e.g. to Mars). The mission also offers unprecedented possibilities to study the Sun's corona and the inner heliosphere. Any CME with an angular extent of 60 degrees or larger will be monitored in-situ and remotely, leading to a better understanding of a CME'€™s morphology and its spatial as well as temporal distribution. Our view of the Sun's corona will change from 2D to 3D, giving rise to new unexpected discoveries. Additionally, the gravity assist with Venus presents a case study on its own and allows for analysis of Venus's magnetic field through multipoint in-situ measurements.\\
This mission will not only provide information vital to protect the health, economy, and modern technology of the human society. It will also bring space weather monitoring to a higher level, and propel advancement in the scientific community.


\bigskip
\bigskip
\bigskip

\begin{acknowledgements}

The authors acknowledge the contribution of Daniel Schmid, Ioanna Patsou, Joseph Nguyen, and Markus Suter, during the Alpbach summer school and are very grateful for the expertise of Joe Zender and Rumi Nakamura, who proved to be invaluable during the Alpbach summer school.\\
Furthermore the authors are greatly indebted for the advice and efforts from Peter Falkner, Christian Erd, and Markus Hallmann, as with their help this mission developed to its current shape.\\
Nikolaos Perakis and Francesco Checco Gini offered very good support regarding orbital calculations, and Viktor Andersson for helped with thermal aspects.\\
The reviewing of the article by G\"unther Reitz and Luciano Rodriguez proved valuable and is highly appreciated.\\
Aditionally, the authors are grateful for all the organisational effort of Michaela Gitsch, who is of great help to young scientists by introducing them into the (new) field of space science and engineering.\\
Finally, the authors acknowledge the financial support by the European Space Agency (ESA), the Aeronautics and Space Agency of Austria FFG, the International Space Science Institute (ISSI), and Austrospace, the association of Austrian space industries and research institutions, all of them making the Alpbach summer school financially possible. In addition, all participants of the summer school were funded by organisations from their respective member states.\\
The editor thanks Norma B. Crosby and an anonymous referee for their assistance in evaluating this paper.

\end{acknowledgements}

\end{document}